\documentstyle[12pt]{article}
\def\nabstar#1{\nabla\kern-0.5pt\smash{\raise 4.5pt\hbox{$\ast$}}
               \kern-4.5pt_{#1}}

\def\drvstar#1{\partial\kern-0.5pt\smash{\raise 4.5pt\hbox{$\ast$}}
               \kern-5.0pt_{#1}}

\def\newline{\relax\ifhmode\null\hfil\break\else\nonhmodeerr@\newline\fi}
\def\frac#1#2{{#1\over#2}}
\def\text#1{{\hbox{\rm #1}}}
\def\flushpar{{\par \noindent}}

\newcommand{\beq}{\begin{equation}}
\newcommand{\eeq}{\end{equation}}
\newcommand{\bea}{\begin{eqnarray}}
\newcommand{\eea}{\end{eqnarray}}

\def\BE{\begin{equation}}
\def\EE{\end{equation}}
\def\BA{\begin{eqnarray}}
\def\EA{\end{eqnarray}}
\def\BAN{\begin{eqnarray*}}
\def\EAN{\end{eqnarray*}}

\def\nn{\nonumber\\}

\def\gm5{\gamma_5}

\input epsf.sty
\newdimen\psfigsize
\def\psfigure#1 #2 #3 #4 #5{
    \begin{figure}[tbh]
      \begin{center}
      \vbox{
        \null\vskip-0.2in\hskip#2
        \epsfxsize=#1
        \epsfbox{#4}
        \vskip -0.3in
        \caption {#5 \label{#3}}
        \vskip 0.0 true in plus 0.3 true in
      }
      \end{center}
   \end{figure}
}
\begin{document}
\thispagestyle{empty}
\begin{flushright}
NTUTH-00-232 \\
August 2000
\end{flushright}
\bigskip\bigskip\bigskip
\vskip 2.5truecm
\begin{center}
{\LARGE {Can nonlocal Dirac operators be topologically proper ?}}
\end{center}
\vskip 1.0truecm
\centerline{Ting-Wai Chiu}
\vskip5mm
\centerline{Department of Physics, National Taiwan University}
\centerline{Taipei, Taiwan 106, Republic of China.}
\centerline{\it E-mail : twchiu@phys.ntu.edu.tw}
\vskip 2cm
\bigskip \nopagebreak \begin{abstract}
\noindent

By examining the analyticity of a sequence of
topologically-proper lattice Dirac operators,
we show that they tend to a nonlocal Dirac operator.
This implies that a nonlocal lattice Dirac operator
can have exact zero modes satisfying the Atiyah-Singer
index theorem, in gauge backgrounds with nonzero topological
charge.



\end{abstract}
\vskip 1.5cm

\newpage\setcounter{page}1

It is well known that all fundamental interactions are local.
Therefore, it is natural to require that any lattice
( nonperturbative ) formulation of these theories is local in
the continuum limit. Although it is straightforward to formulate
bosonic fields on the lattice with ultralocal operators,
it is nontrivial to formulate the Dirac fermion field on the
lattice, which can retain all its vital features in the continuum.
In general, a decent lattice Dirac operator $ D $ is required to satisfy
the following conditions :
\begin{description}
\item[(i)] $ D $ is local. \newline
         ( $ || D(x,y) || \le c \exp( - | x - y | / l ) $
          with $ l \sim a $; or $ D(x,y) = 0 $ for $ |x-y| $ larger
          than a finite distance. )
\item[(ii)] In the free fermion limit, $ D $ is free of species doublings.
             \newline
         ( The free fermion propagator $ D^{-1}(p) $ has only one simple
           pole at the origin $ p = 0 $ in the Brillouin zone. )
\item[(iii)] In the free fermion limit, $ D $ has the correct continuum
             behavior. \newline
         ( In the limit $ a \to 0 $, $ D(p) \sim i \gamma_\mu p_\mu $
           around $ p = 0 $. )
\item[(iv)] $ D $ is $\gm5$-hermitian.
            ( $ D^{\dagger} = \gm5 D \gm5 $. )
\item[(v)] $ D $ satisfies the Ginsparg-Wilson relation \cite{gwr}. \newline
           ($ D \gm5 + \gm5 D = 2 a  D \gamma_5 R D $, where $ R $ is a
          positive definite Hermitian operator which commutes with $ \gm5 $.)
\end{description}

However, there exists {\it no} proof that {\it any} lattice Dirac operator
satisfying these five conditions must have exact zero modes satisfying
the Atiyah-Singer index theorem \cite{AS_thm}, in background gauge fields
with nonzero topological charge. Though one has no doubts that
conditions {\bf (ii)-(v)} are necessary requirements for any topologically
proper $ D $, it is unclear how the locality of $ D $
could play any role in the index of $ D $.
For example, consider the chirally symmetric Dirac operator
\bea
\label{eq:Dc}
D_c = a^{-1} \frac{ \sqrt{ H_w^2} + \gm5 H_w }
                  { \sqrt{ H_w^2} - \gm5 H_w }
\eea
where
\bea
\label{eq:Hw}
H_w &=& \gm5 \left( \gamma_\mu t_\mu + W - \frac{1}{2a} \right) \ , \\
\label{eq:tmu}
t_\mu (x,y) &=& \frac{1}{2a} \ [   U_{\mu}(x) \delta_{x+\hat\mu,y}
                       - U_{\mu}^{\dagger}(y) \delta_{x-\hat\mu,y} ] \ , \\
\label{eq:wilson}
W(x,y) &=&  \frac{1}{2a} \sum_\mu \left[ 2 \delta_{x,y}
                     - U_{\mu}(x) \delta_{x+\hat\mu,y}
                     - U_{\mu}^{\dagger}(y) \delta_{x-\hat\mu,y} \right] \ ,
\eea
\BAN
\gamma_\mu = \left( \begin{array}{cc}
                            0                &  \sigma_\mu    \\
                    \sigma_\mu^{\dagger}     &       0
                    \end{array}  \right) \ ,
\EAN
and
\BAN
\sigma_\mu \sigma_\nu^{\dagger} + \sigma_{\nu} \sigma_\mu^{\dagger} =
2 \delta_{\mu \nu} \ .
\EAN
The Dirac and color indices have been suppressed in
(\ref{eq:Dc})-(\ref{eq:wilson}).
It is obvious that $ D_c $ is topologically-proper and nonlocal,
in the continuum limit. However, one can transform this nonlocal $ D_c $
into a local $ D $ which has exactly the same index of $ D_c $ \cite{twc98:6a},
\bea
\label{eq:tit}
D = D_c ( 1 + a D_c )^{-1} \ .
\eea
Substituting (\ref{eq:Dc}) into (\ref{eq:tit}), we recover
the Overlap Dirac operator \cite{hn97:7,rn95}
\bea
\label{eq:hn}
D = \frac{1}{2} \left( 1 + \gm5 \frac{ H_w }{ \sqrt{H_w^2} } \right) \ ,
\eea
which is local if the gauge background is sufficiently smooth at
the scale of the lattice spacing \cite{ml98:8a,hn99:11}.
Thus, the crucial point is the existence of a topologically proper
Dirac operator, whether it is a nonlocal $ D_c $
or a local $ D $ really does {\it not} matter at all. We can always
transform a nonlocal $ D_c $ into a well-defined and local $ D $
which has the same index of $ D_c $.
From this point of view, the locality of $ D $ does not seem to
play any role in the index of $ D $.

In this paper, we provide a nontrivial example of $ D $ which
is {\it nonlocal} and {\it topologically-proper}. Unlike the chirally
symmetric $ D_c $ (\ref{eq:Dc}) which has poles in the nontrivial sectors,
as a consequence of the chirality sum rule \cite{twc98:4},
the present example breaks the chiral symmetry according to the GW
relation, thus it is well-defined in the nontrivial sectors.

Consider the sequence of topologically-proper lattice Dirac
operators as proposed by Fujikawa \cite{fuji00:4}
\bea
\label{eq:Dk}
D = a^{-1} \left( \frac{1}{2} \right)^{1/(2k+1)}
    \gm5 \left( \gm5 + \frac{H}{\sqrt{H^2}} \right )^{1/(2k+1)},
    \hspace{4mm} k=0, 1, 2,\cdots
\eea
where
\bea
\label{eq:Hk}
H &=& \gm5 [  (-1)^{k} (\gamma_\mu t_\mu)^{2k+1}
             + W^{2k+1} - ( m_0 a^{-1} )^{2k+1}   ],
\hspace{4mm} 0 < m_0 < 2 , \
\eea
and $ t_\mu $ and $ W $ are defined in (\ref{eq:tmu}) and (\ref{eq:wilson}).
Note that the $(2k+1)$-th real root of the Hermitian operator
$ ( \gm5 + H/\sqrt{H^2} ) $ is well-defined.
Recall that any Hermitian operator $ h $ can be
diagonalized by a unitary transformation
\BAN
h = U \Lambda U^{\dagger} \ ,
\EAN
where $ \Lambda $ denotes the diagonal matrix
$ \{ \lambda_1, \lambda_2, \cdots \} $ consisting of the real eigenvalues
of $ h $, and $ U $ the unitary matrix formed by columns of
corresponding normalized eigenvectors $ \{ u_1, u_2, \cdots \} $.
Then any integer power of $ h $ can be written as
\BAN
h^{\alpha}=U \Lambda^{\alpha} U^{\dagger}, \hspace{4mm} \alpha = \mbox{integer}.
\EAN
Thus, the $(2k+1)$-th real root of $ h $ can be defined by
\bea
\label{eq:h_root}
h^{1/(2k+1)} = U \Lambda^{1/(2k+1)} U^{\dagger} \ ,
\eea
where
\bea
[\Lambda^{1/(2k+1)}]_{ij} = \delta_{ij} \
        \mbox{sign}(\lambda_i) \ | \lambda_i |^{1/(2k+1)} \ .
\eea
It is obvious that $ h^{1/(2k+1)} $ defined in (\ref{eq:h_root})
satisfies the relation
\BAN
 ( h^{1/(2k+1)} )^{2k+1} = h \ .
\EAN

For $ k=0 $, $ D $ reduces to the Overlap-Dirac operator (\ref{eq:hn}).
In general, $ D $ satisfies the Ginsparg-Wilson relation \cite{gwr}
\bea
\label{eq:gwr}
D \gm5 + \gm5 D = 2 a D R \gm5 D \
\eea
with
\bea
\label{eq:fuji_R}
R = ( a \gm5 D )^{2k}, \hspace{4mm} k=0, 1, 2, \cdots
\eea
where $ R $ is Hermitian and commutes with $ \gm5 $.

The general properties of $ D $ have been derived in ref. \cite{twc00:5}.
In particular, in the naive continuum limit,
the index of $ D $ is independent of $ k $,
\beq
\label{eq:index_sym}
\mbox{index}[D(m_0)] = \left\{  \begin{array}{ll}
\frac{ (-1)^{n+1} (d-1)! }{ (d-n)! \ (n-1)! } \ Q \ ,
     & \mbox{ \ $ 2(n-1) < m_0 < 2 n $} \\
$ $  & \mbox{ \ \  \ for $ n=1,\cdots,d $ ; }            \\
 0,  & \mbox{ otherwise.   }  \\
       \end{array}
       \right.
\eeq
where $ d $ ( even integer ) is the dimensionality of the lattice,
and $ Q $ is the topological charge of the gauge background.
For $ m_0 \in ( 0, 2 ) $, the index of $ D $ is equal to $ Q $,
hence $ D $ is topologically proper.

One of the salient features of the sequence of topologically-proper
Dirac operators (\ref{eq:Dk}) is that the amount of
chiral symmetry breaking ( i.e., r.h.s. of (\ref{eq:gwr}) )
decreases as the order $ k $ increases.
However, at finite lattice spacing, the chiral symmetry breaking of $ D $
{\it cannot} be zero even in the limit $ k \to \infty $,
since $ D $ satisfies the GW relation (\ref{eq:gwr}).
The only possibility for $ D $ to break the chiral symmetry in the limit
$ k \to \infty $ is that $ D (p) $ becomes a piecewise
continuous function in the Brillouin zone, with discontinuities somewhere
at $ | p | \simeq a^{-1} $, and $ \gm5 D(p) + D(p) \gm5 = 0 $ for
$ | p | < a^{-1} $. Since such a $ D(p) $ is non-analytic at infinite
number of $ p $, the corresponding $ D(x,y) $ must be nonlocal in the
position space.

If $ D(x,y) $ is nonlocal in the free fermion limit
( i.e., $ U_\mu(x) \to 1 $ ), then it must be nonlocal
in any smooth gauge background. So, it suffices to examine
the non-analyticity of $ D(p) $ in the free fermion limit.

In the free fermion limit, the GW Dirac operator (\ref{eq:Dk}) in momentum
space can be written as
\bea
\label{eq:Dkp}
D(p)   &=&   D_0(p) + i \gamma_\mu D_\mu(p) \ , \\
\label{eq:D0}
D_0(p) &=& a^{-1} \left[ \frac{1}{2} \left( 1 + \frac{u(p)}{N(p)} \right)
                \right]^{\frac{k+1}{2k+1}} \ ,  \\
\label{eq:Dmu}
D_\mu(p) &=& a^{-1} \left[ \frac{1}{2} \left( 1 + \frac{u(p)}{N(p)} \right)
                    \right]^{ \frac{k+1}{2k+1} }
\sqrt{\frac{1}{a^2 t^2(p)} \left( \frac{N(p)-u(p)}{N(p)+u(p)} \right) }
               \sin( p_\mu a ) \ , \nn
\eea
where
\bea
\label{eq:up}
u(p) &=& a^{-(2k+1)}
         \left( \left[ \sum_\mu \ ( 1 - \cos( p_\mu a ) ) \right]^{2k+1}
                             - m_0^{2k+1} \right) \ , \\
\label{eq:t2}
t^2(p) &=& a^{-2} \sum_\mu \sin^2( p_\mu a ) \ , \\
\label{eq:Np}
N(p)  &=& \sqrt{ [t^2(p)]^{2k+1}+ [u(p)]^2 } \ .
\eea
In the following, we set
\bea
 m_0 = \left( \frac{1}{2} \right)^{1/(2k+1)}
\eea
such that $ D(p) = i \gamma_\mu p_\mu $ in the naive continuum limit.

The chiral symmetry breaking in $ D(p) $ is due to its
scalar component $ D_0(p) $. For $ a | p | \ll 1 $,
\bea
D_0 ( p ) &=& a^{-1} ( a^2 p^2 )^{k+1} + \mbox{higher order terms} \ , \\
D_\mu ( p ) &=& p_\mu ( 1 + O(a^2 p_\mu^2 ) ) \ .
\eea
Thus, at small momenta or in the naive continuum limit, the chiral
symmetry breaking of $ D_0(p) $ decreases as the order $ k $ increases.
On the other hand, at the $ 2^d - 1 $ corners of the Brillouin zone,
$ D_0( p )= a^{-1} $ for all $ k $. Therefore, $ D_0(p) $ must undergo
a transition from zero to $ a^{-1} $ as any momentum component
$ | p_\mu | $ is increased from zero to $ \pi/a $. The region of
transition is essentially confined between two closed
concentric hypersurfaces inside the Brillouin zone.
As the order $ k $ gets higher, the region of transition
becomes thinner and the slope of the transition gets steeper.
In the limit $ k \to \infty $, the transition occurs only at
a few hypersurfaces ( i.e., the transition is composed of several
step functions ), thus renders $ D_0(p) $ non-analytic and
$ D_0 (x) $ nonlocal.

It is instructive to work out the behavior of $ D_0(p) $ in two
dimensions (  i.e., $ p_3 = p_4 = 0 $ ).
In the limit $ k \to \infty $, (\ref{eq:D0}) becomes
\bea
\label{eq:D01}
D_0(p) = a^{-1} \sqrt{ \frac{1}{2}
\left( 1 + \lim_{k \to \infty} \frac{u(p)}{N(p)} \right) } \ .
\eea
On the $ ( p_1, p_2 ) $ plane, (\ref{eq:up})-(\ref{eq:Np}) become
\bea
\label{eq:up2}
u(p)  &=& a^{-(2k+1)} \left[ ( 2 - c_1 - c_2 )^{2k+1}
                                 - \frac{1}{2} \right] \ , \\
\label{eq:t22}
t^2(p) &=& a^{-2} ( s_1^2 + s_2^2 ) \ , \\
\label{eq:Np2}
N(p)  &=&  \sqrt{ [t^2(p) ]^{2k+1} + [u(p)]^2   } \ .
\eea
where
\bea
c_\mu = \cos( p_\mu a ) \ ,  \hspace{4mm}  s_\mu = \sin( p_\mu a ) \ .
\eea
Note that the factor $ a^{-(2k+1)} $ in (\ref{eq:up2}) and (\ref{eq:Np2})
is cancelled in the ratio $ u(p)/N(p) $.

First we consider $ ( p_1, p_2 ) $ inside the square  ( see Fig. 1 )
bounded by the four edges : $ | p_1 | + | p_2 | = \pi/2a $.
It satisfies
\BAN
  | p_1 | + | p_2 | &<& \pi/2a  \ , \\
   s_1^2  + s_2^2   &<& 1 \ , \\
    2 - c_1 - c_2   &<& 1 \ .
\EAN
Hence
\BAN
\lim_{k \to \infty } \frac{[t^2(p)]^{2k+1}}{[u(p)]^2}
                    =\frac{0}{(-1/2)^2}=0 \ , \\
\lim_{k \to \infty } \frac{u(p)}{N(p)} = - 1 \ .
\EAN
From (\ref{eq:D01}), we have
\bea
\lim_{k \to \infty } D_0(p) = 0 \ , \ \mbox{ for } \
 | p_1 | + | p_2 | < \frac{\pi}{2a} \ .
\eea

Next we consider $ ( p_1, p_2 ) $ lying on the edge of the square,
except its four vertices $ \{ ( \frac{\pi}{2a}, 0 ), ( 0, \frac{\pi}{2a} ),
( -\frac{\pi}{2a}, 0 ), ( 0, -\frac{\pi}{2a} ) \} $. It satisfies
\BAN
 | p_1 | + | p_2 | &=& \pi/2a  \ , \\
    s_1^2 + s_2^2  &=& 1 \ , \\
    2 - c_1 - c_2  &<& 1 \ .
\EAN
Thus
\BAN
\lim_{k \to \infty } \frac{[t^2(p)]^{2k+1}}{[u(p)]^2}
                    = \frac{1}{(-1/2)^2} = 4 \ , \\
 \lim_{k \to \infty} \frac{u(p)}{N(p)}
                    = -\frac{1}{\sqrt{5}} \ .
\EAN
From (\ref{eq:D01}), we have
\bea
\lim_{k \to \infty } D_0(p) =
a^{-1} \sqrt{ \frac{1}{2} \left( 1 - \frac{1}{\sqrt{5}} \right) } \ ,
\ \mbox{for} \ | p_1 |+| p_2 |=\frac{\pi}{2a}, \ |p_i| \ne \frac{\pi}{2a} \ .
\eea

Next we consider the four vertices
$ \{ ( \frac{\pi}{2a}, 0 ), ( 0, \frac{\pi}{2a} ),
( -\frac{\pi}{2a}, 0 ), ( 0, -\frac{\pi}{2a} ) \} $ of the square.
Each one of them satisfies
\BAN
  s_1^2  + s_2^2  &=& 1 \ , \\
   2 - c_1 - c_2  &=& 1 \ .
\EAN
Thus
\BAN
 \lim_{k \to \infty} \frac{u(p)}{N(p)} = \frac{1/2}{\sqrt{ 1 + (1/2)^2 } }
                                       = \frac{1}{\sqrt{5}} \ .
\EAN
From (\ref{eq:D01}), we have
\bea
\lim_{k \to \infty } D_0(p) =
a^{-1} \sqrt{\frac{1}{2} \left(1+\frac{1}{\sqrt{5}} \right) } \ ,\ \mbox{for}
\ (p_1, p_2)=\{ ( \pm \frac{\pi}{2a}, 0 ); \ ( 0, \pm \frac{\pi}{2a} ) \} \ .
\eea

Next we consider $ ( p_1, p_2 ) $ lying on the closed curve ( see Fig. 1 ),
\BAN
( 2 - c_1 - c_2 )^2 = s_1^2 + s_2^2 \ ,
\EAN
which circumscribes the square. The equation of this closed
curve can be rewritten as
\bea
\label{eq:cc}
( 1- c_1 )^2 + ( 1- c_2 )^2 + c_1 c_2 = 1 \ .
\eea
Except at the four vertices,
each $ (p_1, p_2 ) $ along this curve satisfies
\BAN
 2 - c_1 - c_2  &>& 1 \ , \\
 s_1^2 + s_2^2  &>& 1 \ .
\EAN
Thus,
\BAN
\lim_{k \to \infty } \frac{[t^2(p)]^{2k+1}}{[u(p)]^2} &=& 1 \ , \\
\lim_{k \to \infty } \frac{u(p)}{N(p)} &=&  \frac{1}{\sqrt{2}} \ .
\EAN
From (\ref{eq:D01}), we have
\BAN
\lim_{k \to \infty } D_0(p) =
a^{-1} \frac{\sqrt{ 2 + \sqrt{2} }}{2} \ , \
\mbox{ for } \ (p_1, p_2) \ \mbox{satisfying} \ (\ref{eq:cc}) \ \mbox{and} \
                | p_1 |, |p_2| \ne \frac{\pi}{2a} \ .
\EAN

Next we consider $ ( p_1, p_2 ) $ inside the region bounded by the closed
curve and the four edges of the square,
except the four vertices ( see Fig. 1 ).
It satisfies
\BAN
          ( 2 - c_1 - c_2 )^2 &<& s_1^2 + s_2^2 \ , \\
           s_1^2 + s_2^2      &>& 1 \ .
\EAN
Thus
\BAN
\lim_{k \to \infty } \frac{[u(p)]^2}{[t^2(p)]^{2k+1}} = 0  \ , \\
\lim_{k \to \infty } \frac{u(p)}{N(p)} = 0 \ .
\EAN
From (\ref{eq:D01}), we have
\BAN
\lim_{k \to \infty } D_0(p) =
a^{-1} \frac{1}{\sqrt{2}} \ ,
\mbox{ for }   ( 1- c_1 )^2 + ( 1-c_2 )^2 + c_1 c_2 < 1, \
                | p_1 |+| p_2 | > \frac{\pi}{2a}.
\EAN

Finally, we consider $ (p_1, p_2) $ lying outside the closed curve
up to the boundary of the Brillouin zone ( see Fig. 1 ).
It satisfies
\BAN
            ( 2 - c_1 - c_2 )^2 > s_1^2 + s_2^2 > 1 \ .
\EAN
Thus
\BAN
\lim_{k \to \infty} \frac{ [t^2(p)]^{2k+1} }{ [u(p)]^2 }
= \lim_{k \to \infty} \left( \frac{ s_1^2 + s_2^2 }
                  { ( 2 - c_1 - c_2 )^2 } \right)^{2k+1}
                 = 0 \ .
\EAN
Then
\BAN
\lim_{k \to \infty } \frac{u(p)}{N(p)} = 1 \ .
\EAN
From (\ref{eq:D01}), we have
\BAN
\lim_{k \to \infty } D_0(p) = a^{-1} \ , \  \mbox{ for } \
( 1- c_1 )^2 + ( 1- c_2 )^2 + c_1 c_2 > 1 \ .
\EAN

To summarize,
\beq
\label{eq:D0p_2}
\lim_{k \to \infty } D_0(p) = \left\{  \begin{array}{ll}
  0 \ , & | p_1 | + | p_2 | < \frac{\pi}{2a} \ ; \\
a^{-1}  \left(  \frac{1}{2} - \frac{\sqrt{5}}{10} \right)^{\frac{1}{2}} \ ,
      & | p_1 |+| p_2 |=\frac{\pi}{2a},
        \ |p_1|\ne \frac{\pi}{2a}, \ |p_2| \ne \frac{\pi}{2a} \ ; \\
a^{-1}  \left(  \frac{1}{2} + \frac{\sqrt{5}}{10} \right)^{\frac{1}{2}} \ ,
      & (p_1, p_2)=\{ (\pm\frac{\pi}{2a},0), (0,\pm\frac{\pi}{2a} ) \} \ ; \\
a^{-1} \frac{\sqrt{2}}{2} \ ,
      & ( 1- c_1 )^2 + ( 1- c_2 )^2 + c_1 c_2 < 1, \
         | p_1 |+| p_2 | > \frac{\pi}{2a} \ ; \\
a^{-1} \frac{\sqrt{2+\sqrt{2}}}{2} \ ,
      & ( 1- c_1 )^2 + ( 1- c_2 )^2 + c_1 c_2 = 1 \ ; \\
a^{-1} \ ,   & ( 1- c_1 )^2 + ( 1- c_2 )^2 + c_1 c_2 > 1 \ .
       \end{array}
       \right.
\eeq

In short, $ D_0(p) $ in the Brillouin zone is essentially segregated
into three piecewise constant parts.
It is zero inside the square, $ \sqrt{2}/2a $ inside the region bounded
by the closed curve and the four edges of the inscribed square,
and $ a^{-1} $ outside the closed curve.

The discontinuities of $ D_0(p) $ at the four edges of the inscribed
square ( $ | p_1 |+| p_2 | = \frac{\pi}{2a} $ ) is
a jump from zero to $ \sqrt{2} / 2 a $, while
those at the closed curve [ $ ( 1- c_1 )^2 + ( 1- c_2 )^2 + c_1 c_2 = 1 $ ]
is another jump from $ \sqrt{2} / 2 a $
to $ a^{-1} $. At the four vertices of the inscribed square,
these two step functions merge together.

\psfigure 5.0in -0.2in {fig:curve} {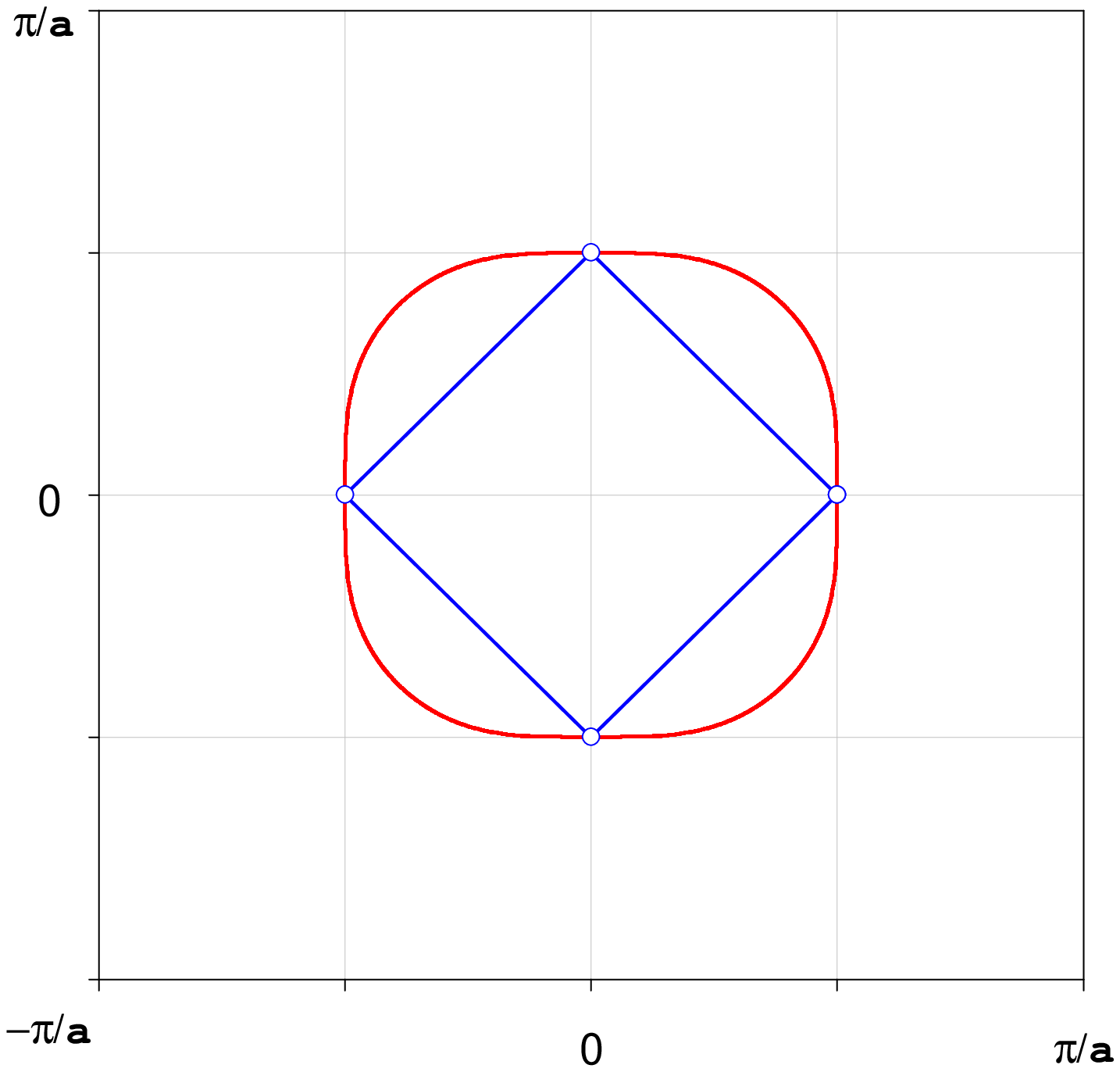} {
The discontinuities of $ D_0(p) $ in the two-dimensional Brillouin zone.
$ D_0(p) $ is zero for $ | p_1 | + | p_2 | < \pi/2a $, i.e.,
inside the inscribed square. Then $ D_0(p) $ jumps from zero to
$ \sqrt{2}/2a $ at the four edges of the square, and remains $ \sqrt{2}/2a $
for all $ ( p_1, p_2 ) $ inside the region bounded by the four
edges of the square and the closed curve
[ $ ( 1- \cos(p_1) )^2 + ( 1- \cos(p_2) )^2 + \cos(p_1) \cos(p_2) = 1 $ ].
Finally it jumps from $ \sqrt{2}/2a $ to $ a^{-1} $ at the closed curve,
and remains $ a^{-1} $ for all $ ( p_1, p_2 ) $ outside the closed
curve. }

By repeating above analysis, one can obtain the
discontinuities of $ D_\mu(p) $ (\ref{eq:Dmu}) in the limit $ k \to \infty $.
Their locations are the same as those of $ D_0(p) $, at the closed
curve as well as at the four edges of the inscribed square.

Along the diagonals ( $ | p_1 | = | p_2 | $ ) of the Brillouin zone,
\beq
\label{eq:D1p_2}
\lim_{k \to \infty } D_\mu(p) =  \left\{  \begin{array}{ll}
 p_\mu \ , & | p_1 | = | p_2 | < \frac{\pi}{4a} \ ; \\
\frac{1}{2a} \sqrt{1 + \frac{1}{\sqrt{5}} } \ ,
     & | p_1 | = | p_2 |=\frac{\pi}{4a} \ ; \\
\frac{1}{2a} \ ,
      & \frac{\pi}{4a} < | p_1 | = | p_2 | < \cos^{-1}(\frac{1}{3}) \ ; \\
\frac{1}{2a} \sqrt{ 1 - \frac{1}{\sqrt{2}} } \ ,
      &  | p_1 | = | p_2 | = \cos^{-1}(\frac{1}{3}) \ ; \\
0 \ , &  | p_1 | = | p_2 | > \cos^{-1}(\frac{1}{3}) \ .
       \end{array}
       \right.
\eeq

In general, in the limit $ k \to \infty $,
$ D_\mu(p) = p_\mu $ for $ | p | < \pi/4a  $.
As $ | p | $ is increased further, $ | D_\mu(p) | $ bends down and its
slope becomes less than one. Then it undergoes a discontinuous transition
at the edges of the square ( $ | p_1 | + | p_2 | = \pi/2a $ ),
finally it drops to zero discontinuously at the closed curve
[ $ ( 1- c_1 )^2 + ( 1- c_2 )^2 + c_1 c_2 = 1 $ ],
and remains zero up to the boundary of the Brillouin zone.

It is straightforward to repeat above analysis in
four ( and higher ) dimensions\footnote{In $ d $ dimensions, the
discontinuities of $ D(p) $ in the limit $ k \to \infty $ are located
at two concentric closed hypersurfaces. The inscribed one is
\BAN
\sum_\mu s_\mu^2 = 1 \ ,
\EAN
while the circumscribing one is
\BAN
\frac{d(d-1)}{2} + \sum_\mu c_\mu ( c_\mu  - d )
                 + \sum_{\mu > \nu} c_\mu c_\nu  = 0 \ ,
\EAN
where $ c_\mu = \cos( p_\mu a ) $ and $ s_\mu = \sin( p_\mu a ) $.
Their intersections are located at :
\BAN
d - \sum_\mu c_\mu = \sum_\mu s_\mu^2 = 1 \ .
\EAN
},
however, it is unnecessary for
our present purpose. Since the infinite number of discontinuities
of $ D(p) $ in two dimensions implies that $ D(p) $
also has infinite number of discontinuities in four dimensions.
This renders $ D(x,y) $ nonlocal in four ( and higher ) dimensions
as well as in two dimensions.

In this paper, we have examined the analyticity of a sequence of
topologically proper lattice Dirac operators (\ref{eq:Dk}).
We have shown that in the limit $ k \to \infty $,
$ D(p) $ has infinite number of discontinuities. This implies
that $ D(x,y) $ is nonlocal in the limit $ k \to \infty $.
On the other hand, $ D(x,y) $ is local at the zeroth order ( $ k = 0 $ ),
for gauge backgrounds which are sufficiently smooth at the scale of the
lattice spacing \cite{ml98:8a,hn99:11}.
Since the index of $ D $ (\ref{eq:Dk}) is independent of the order $ k $,
this suggests that the locality of a lattice Dirac operator is
irrelevant to its index. In other words, if a lattice Dirac operator
satisfies the four conditions {\bf (ii)-(v)} but always has zero index
\cite{twc99:11}, then the cause may {\it not} be due to its nonlocality.

\eject

\flushpar
{\bf Acknowledgement }
\bigskip

\noindent

This work was motivated after a visit to The State University of New Jersey,
Rutgers. I am grateful to Herbert Neuberger for his kind hospitality,
as well as many helpful discussions. I also enjoyed the discussions together
with Rajamani Narayanan and Federico Berruto. I would like to thank Jim Ball
and Carleton DeTar for their kind hospitality and discussions during my visit
to The University of Utah. This work was supported by the National Science
Council, Republic of China, under the grant number NSC89-2112-M002-079.

\bigskip
\bigskip


\end{document}